\documentclass[10pt,conference]{IEEEtran}
\IEEEoverridecommandlockouts
\usepackage{cite}
\usepackage{amsmath,amssymb,amsfonts}
\usepackage{algorithmic}
\usepackage{graphicx}
\usepackage{textcomp}
\usepackage{xcolor}
\usepackage{booktabs} 
\usepackage{listings}
\usepackage{multirow}
\usepackage{soul}
\usepackage{xspace}
\usepackage{graphicx}
\usepackage{ifthen}
\usepackage[normalem]{ulem} 
\usepackage{xcolor,colortbl}
\usepackage{hyperref}

\newboolean{showedits}
\setboolean{showedits}{true} 
\ifthenelse{\boolean{showedits}}
{
	\newcommand{\del}[1]{\textcolor{red}{\sout{#1}}} 
	\newcommand{\nbe}[3]{
		{\colorbox{#3}{\bfseries\sffamily\scriptsize\textcolor{white}{#1}}}
		{\textcolor{#3}{\sf\small$\blacktriangleright$\textit{#2}$\blacktriangleleft$}}}
}{
	\newcommand{\del}[1]{} 
	
	\newcommand{\nbe}[3]{}
}

\newboolean{showcomments}
\setboolean{showcomments}{true} 
\newcommand{\id}[1]{$-$Id: scgPaper.tex 32478 2010-04-29 09:11:32Z oscar $-$}

\ifthenelse{\boolean{showcomments}}
 {
 	\newcommand{\nbc}[3]{
 		{\colorbox{#3}{\bfseries\sffamily\scriptsize\textcolor{white}{#1}}}
		{\textcolor{#3}{\sf\small$\blacktriangleright$\textit{#2}$\blacktriangleleft$}}}
	
 }{
 	\newcommand{\nbc}[3]{}
 	
 }

\usepackage[most]{tcolorbox}
\ifthenelse{\boolean{showedits}}
{
  \newtcolorbox{inserted}{%
       title=Inserted text:,
       colframe=blue,colback=blue!5!white,
       breakable,
       leftrule=0mm, 
       bottomrule=0mm,
       rightrule=0mm,
       toprule=0mm,
       arc=0mm, outer arc=0mm,
       oversize
  }
  \newtcolorbox{deleted}{%
       title=Deleted text:,
       colframe=red,colback=red!5!white,
       breakable,
       leftrule=0mm, 
       bottomrule=0mm,
       rightrule=0mm,
       toprule=0mm,
       arc=0mm, outer arc=0mm,
       oversize
  }
  \newtcolorbox{refactored}{%
       title=Rewritten text:,
       colframe=blue,colback=red!5!white,
       breakable,
       leftrule=0mm, 
       bottomrule=0mm,
       rightrule=0mm,
       toprule=0mm,
       arc=0mm, outer arc=0mm,
       oversize
  }
}{

}
\newboolean{isblinded}
\setboolean{isblinded}{true}
\ifthenelse{\boolean{isblinded}}
{\newcommand\blind[1]{BLINDED\xspace}}
{\newcommand\blind[1]{#1\xspace}}

\newcommand{\commented}[1]{}

\newcommand{\eg}{\emph{e.g.,}\xspace}
\newcommand{\ie}{\emph{i.e.,}\xspace}
\newcommand{\etal}{\emph{et al.}\xspace}

\definecolor{source}{gray}{0.95}
\usepackage[T1]{fontenc} 

\newcommand{\boxit}[1]{\vspace{0.2cm}
\noindent
\fbox{
\begin{minipage}{8.3cm}
\normalsize \emph{#1} 
\end{minipage}
}
}

\lstdefinelanguage{Java}{
  tabsize=4
}[keywords,comments,strings]

\definecolor{source}{gray}{0.95}
\definecolor{highlight}{gray}{0.9}

\lstnewenvironment{codesnippet}{%
	\lstset{%
		frame=single,
		framerule=0pt,
		mathescape=false
	}
}{}

\newcommand{\SO}{Stack Overflow\xspace}

\begin{document}
\title{Dazed and Confused: What's Wrong with Crypto Libraries?}
%
%

\author{\IEEEauthorblockN{Mohammadreza Hazhirpasand}
\IEEEauthorblockA{University of Bern\\
Bern, Switzerland}
\and
\IEEEauthorblockN{Oscar Nierstrasz}
\IEEEauthorblockA{University of Bern\\
Bern, Switzerland}
\and
\IEEEauthorblockN{Mohammad Ghafari}
\IEEEauthorblockA{University of Auckland\\
Auckland, New Zealand}
}

\maketitle              
\begin{abstract}
Recent studies have shown that developers have difficulties in using cryptographic APIs, which often led to security flaws.
We are interested to tackle this matter by looking into what types of problems exist in various crypto libraries.
We manually studied 500 posts on \SO associated with 20 popular crypto libraries.
We realized there are 10 themes in the discussions. 
Interestingly, there were only two questions related to attacks against cryptography. 
There were 63 discussions in which developers had interoperability issues when working with more than a crypto library. 
The majority of posts (112) were about encryption/decryption problems and 111 were about installation/compilation issues of crypto libraries. 
Overall, we realize that the crypto libraries are frequently involved in more than five themes of discussions. 
We believe the current initial findings can help team leaders and experienced developers to correctly guide the team members in the domain of cryptography. 
Moreover, future research should investigate the similarity of problems at the API level among popular crypto libraries.

\end{abstract}

\begin{IEEEkeywords}
Security, Cryptography, Crypto library
\end{IEEEkeywords}
\section{Introduction}
\label{sec:intro}
The literature review shows that often developers do not securely use cryptographic (crypto) APIs.
Hazhirpasand \etal analyzed 2\,324 Java open-source projects in which Java Cryptography Architecture APIs were used.
The authors found that more than 72\% of the projects suffer from at least one cryptographic misuse~\cite{hazhirpasand2019impact}.
Furthermore, they realized using such APIs over a period of time is not an indicative factor to developer performance. 
Rahaman \etal introduced CryptoGuard to detect cryptographic misuses and identified various types of misuses, \eg broken hash, and insecure symmetric and asymmetric algorithms, in 6 181 popular Android applications \cite{rahaman2019cryptoguard}.

Several studies have identified various areas of cryptography that are problematic, and where usability improvement could be advantageous.
Green \etal proposed a guideline including ten principles for creating usable and secure crypto APIs \cite{green2016developers}.
Researchers deduced from 91 954 crypto questions on Stack Overflow that developers mainly have issues in three areas, \ie digital certificates, programming issues, and passwords/hashes, resulting from a distinct lack of knowledge of fundamental concepts and the complexity of crypto libraries \cite{hazhirpasand2021hurdles}.
Lazar \etal studied 269 common vulnerabilities and exposures (CVEs) reports, and devised four main groups of crypto vulnerabilities, \ie plaintext disclosure, man-in-the-middle attacks, brute-force attacks, and side-channel attacks \cite{lazar2014does}.
Nevertheless, the previous work studied the usabilities of crypto APIs and did not study the technical aspect of such APIs.

To advance previous work,
we are interested in shedding light on the following research question: \emph{what technical difficulties are prevalent among crypto libraries?}
We conducted a manual thematic analysis of 500 posts on \SO associated with 20 crypto libraries.
According to the question and answer body in the discussions, we extracted 10 themes.
Most developers (\ie 112 posts) asked how to encrypt/decrypt a piece of text or a file. 
The concern of security attacks against cryptography was observed only in two posts.
We observe (\ie 111 posts) that installation, compilation, and working with different versions of a crypto library form a barrier for developers to start with cryptography.
In summary, our findings indicate that there are more than five themes of problems that involve the majority of the analyzed libraries and can provide assistance for professionals to prepare their team members by addressing problematic areas of crypto libraries beforehand.
In future work, we plan to explore how similar APIs in popular crypto libraries are misunderstood and whether APIs whose complexity is higher than others create more problems or not.

The remainder of this paper is structured as follows.
In \autoref{sec:meth}, we explain the methodology of this study. 
We discuss our findings in \autoref{sec:discussion} and point out potential threats to the validity of this work in \autoref{sec:threat}.
We discuss related work in \autoref{sec:related}, and we conclude this paper in \autoref{sec:conclusion}.
\section{Methodology}
\label{sec:meth}
In the following, we explain the objectives of this study as well as the methodology used for data collection and analysis.

\subsubsection{Objective}
\label{sub:obj}
In this study, we pose the following research question ``what technical difficulties are prevalent among crypto libraries?'' to tackle the underlying reasons why developers' performance varies in using crypto APIs.

The objectives of this research are listed in the following: 
\begin{itemize}
\item Finding prevalent themes of technical challenges in using crypto libraries helps library designers to improve the design of APIs.
\item Underlying factors can assist team leaders to be aware of areas where the developers might encounter difficulties in using cryptography.
\end{itemize}

\subsubsection{Selecting crypto libraries}
We aim at studying posts associated with popular crypto libraries on \SO. We assumed that discussions related to crypto libraries contain the name of the library as a tag. Hence, we selected the “cryptography” tag, \ie base tag, to observe what other tags were used together with the base tag. 
We used Stack Exchange Data Explorer to run a query in order to fetch tags that appeared together with cryptography.\footnote{https://data.stackexchange.com/}
We realized that there are 2 184 tags, \ie candidate tags.
The two authors of this paper separately checked each of the candidate tags.  
Each of the reviewers selected the ones that are crypto libraries. 
They used the internet to explore a tag in which they had a lack of certainty.
Then, they cross-checked the choices and discussed the tags.
They arrived at the conclusion that tags that do not represent a crypto library or only provide a particular, limited service in cryptography (\eg hashing) should not be considered.
As a result, there were 6 tags that were eliminated from the list, namely rsacryptoserviceprovider, aescryptoserviceprovider, rijndaelmanaged, bcrypt, javax.crypto, and hashlib.
The aforementioned tags are either a crypto class, namespace, or a dedicated module only for hashing.
Ultimately, they agreed on a list of 20 crypto libraries, illustrated in \autoref{tab:cryptolib}.

\begin{table}[]
\footnotesize	
\centering
\caption {The selected crypto libraries and their associated number of posts on \SO } \label{tab:cryptolib} 
\begin{tabular}{ll|ll}
\hline
\textbf{Tag name} & \textbf{\# of posts} & \textbf{Tag name} & \textbf{\# of posts} \\ \hline
OpenSSL           & 14 254               & libsodium         & 272                  \\
Bouncy Castle      & 2 799                & M2Crypto          & 263                  \\
CryptoJS          & 1 286                & Web Crypto API     & 215                  \\
mcrypt            & 924                  & JCA               & 200                  \\
PyCrypto          & 842                  & CommonCrypto      & 199                  \\
phpseclib         & 796                  & node-crypto       & 170                  \\
Crypto++          & 713                  & Botan             & 117                  \\
CryptoAPI         & 584                  & Spongy Castle      & 115                  \\
pyOpenSSL         & 436                  & SJCL              & 77                   \\
Jasypt            & 336                  & wolfSSL           & 50                   \\ \hline
\end{tabular}
\end{table}

\subsubsection{Crypto libraries}
The selected crypto libraries are all widely used in practice and have been examined in research projects. 
For instance, six of the selected libraries, \ie OpenSSL, libsodium, Bouncy Castle, SJCL, Crypto-JS, and PyCrypto, were studied for finding usability issues \cite{patnaik2019usability}.
MCrypt is the successor to the Unix crypt command, which supports modern encryption algorithms. \footnote{\url{http://mcrypt.sourceforge.net}}
The phpseclib library offers pure-PHP implementations of SSH2, SFTP, RSA, DSA, and many other algorithms. \footnote{\url{https://github.com/phpseclib/phpseclib}}
Crypto++ and Botan are both C++ crypto libraries that support a wide range of crypto algorithms and security protocols. \footnote{\url{https://www.cryptopp.com}} 
\footnote{\url{https://botan.randombit.net}}
The Microsoft CryptoAPI interface enables developers to employ authentication, encoding, and encryption to Windows-based applications. \footnote{\url{https://docs.microsoft.com/en-us/windows/win32/seccrypto/cryptography-portal}}
Jasypt and Java Cryptography Architecture (JCA) are both intended for Java developers, and the latter is part of the Java security API. \footnote{\url{http://www.jasypt.org/}}\footnote{\url{https://www.oracle.com/java/technologies/javase/javase-tech-security.html}}
The Web Crypto AP is intended to present basic cryptographic operations for web applications and defines cryptographic primitives in a native JavaScript API.
\footnote{\url{https://www.w3.org/TR/WebCryptoAPI/}}
The wolfSSL TLS library is a lightweight, C-language-based library designed for IoT, embedded systems, and smart grids. \footnote{\url{https://www.wolfssl.com/}}
There are also popular OpenSSL wrappers in languages such as node-crypto in Node.js and pyOpenSSL in Python.
There have been numerous studies to investigate the security point of view of aforementioned crypto libraries and their strengths and weaknesses were examined~\cite{cairns2016security} \cite{yarom2017cachebleed} \cite{somorovsky2016systematic}.
However, the security evaluation of these crypto libraries falls outside the scope of this paper.

\subsubsection{Manual investigation}
In total, there are 24 648 posts that contained the selected crypto libraries' tags. 
We computed the required sample size for the population with a confidence level of 95\% and a margin of error of 4.34\%, which results in sampling 500 posts. 
We then equally selected 25 posts from each tag (\ie a crypto library).
We queried the posts containing a crypto library tag,\eg OpenSSL, and set the search criteria to ``recent activity'', so that \SO returns the recent active discussions.
Since we observed questions that are either unanswered or received negative votes, we decided to choose the posts for which the question received at least one upvote and at least one answer.
The list of the selected questions are available online.\footnote{\url{http://crypto-explorer.com/crypto_libs/}}

Thereafter,  we employed thematic analysis, a qualitative research method for finding themes in texts \cite{braun2006using}, to deduce the frequent topics from the chosen posts. 
Since our study is of an exploratory nature, we did not devise a list of themes prior to studying the posts.
Hence, in order to link each post to a suitable theme, two authors of the paper were responsible to separately study the posts and deduce the main issue (\ie theme) of the post.
The reviewers carefully reviewed the title, question body, and answer body of each post. 
Despite the fact that each post may entail several crypto concepts, the reviewers' objective was to find the key issue of each post.
They employed open coding in which a short explanation label was assigned to each post \cite{lewis2015qualitative}.
Each author reiterated the coding phase three times to improve their deduced list of themes.
To evaluate the inter-rater agreement between the two reviewers, we employed Cohen's kappa to assess the agreement level \cite{cohen1960coefficient}.
Deducing the themes from the posts, the reviewers received 68\% Cohen’s Kappa score, which indicates a substantial agreement between the two reviewers.
Finally, the two reviewers compared the two lists and discussed any disagreements.
The two reviewers used different wording for building the list of themes and the total number of themes was not identical.
They re-analyzed the particular posts in multiple sessions where they had different views.
In some scenarios, they realized that one of the reviewers broke down one theme into several sub-themes, which they then merged if necessary. 
Ultimately, they agreed on 10 themes for the analyzed posts.


\section{Results and discussion}
\label{sec:discussion}
\autoref{tab:themes} lists the themes, the associated number of posts in each theme, and a brief summary of what each theme is. 
The highest number of posts is associated with encryption/decryption of a file while the least number of posts is associated with cryptography attacks.
\autoref{tab:alllibs} describes in more detail the number of assigned posts to each theme in the 20 crypto libraries. 
The highlighted cell demonstrates the highest number of posts in each theme compared to other libraries. 
For instance, of 25 analyzed questions in pyOpenSSL, 17 posts were assigned to certificate-related issues.  
In the following, we discuss each of the 10 themes of developer challenges in the 20 crypto libraries.
\begin{table*}
\footnotesize	
\centering
\caption {The deduced themes, number of posts in each theme and associated description} \label{tab:themes} 
\begin{tabular}{lll}
\hline
\textbf{Theme}             & \textbf{\# of posts} & \textbf{Description}                                                              \\ \hline
Encryption/Decryption      & 112                  & Technical problems, \eg modes of encryption, AES or IV, for encryption or decryption of a string  \\
Library installation       & 111                  & Posts related to installation, compilation, and version mismatch                  \\
Certificate-related issues & 74                   & Posts related to SSL, self-signed certificates, PEM, PKCS7, DER                   \\
Library interoperability   & 63                   & Posts related to working with more than a crypto library                          \\
Generate/store keys & 45                   & Posts related to proper methods of loading or generating a crypto key            \\
Hashing                    & 37                   & Posts related to MD5, SHA, HMAC and other hashing algorithms                      \\
Digital signature          & 34                   & Posts related to how to sign or verify a signature                                \\
Sample implementation      & 19                    & Posts where a sample code was requested                                           \\ 
Random number generator    & 3                    & Posts related to generating a true random number                                  \\
Cryptography attacks      & 2                    & Concerns for cryptographic attacks in discussions                                 \\ \hline
\end{tabular}
\end{table*}

\subsubsection{Encryption/Decryption} In this theme, developers struggled with how to conduct file encryption or decryption.
However, the range of sub-problems varies. The first group of challenges is with those who could encrypt a piece of data but the decryption phase was not successful.
For instance, a developer encrypted a string with Spongy Castle and the decryption code was not working due to not employing  AndroidKeyStore for retrieving the private key.
Another observed issue was misusing the \emph{doFinal, init and update} methods in the Cipher API.
A developer missed all the important elements, \ie keys, IV, encoding, and padding,  to perform the decryption process when working with the CryptoJS library.

Another type of discussion was related to the mode of encryptions. For instance, a developer asked for ways of checking the authenticity of an encrypted text and the responses suggested Galois/Counter Mode (GCM) in the AES encryption.
In another discussion, a developer was unsure of the internals of cipher-block chaining (CBC) and why only the first block could be corrupted but the subsequent blocks will be as expected if the initialization vector (IV) is incorrect, whereas in other discussions, either the IV was forgotten in the decryption process or unequal IVs were used.
A developer confused the difference between Electronic codebook (ECB) and CBC, and which one requires the IV for encryption/decryption of a file.
With regards to different modes of encryptions, a common uncertainty was about the correct length of the IV.

One of the prevalent sub-problems was concerning the correct way of encoding/decoding the ciphertext.
For instance, a developer forgot to use UTF-8 to convert plaintext to an array of bits. Other discussions had the same problem of converting the cipher to either hexadecimal or Base64.

There were also some challenges that could not be grouped together, and hence we classified them as miscellaneous. For instance, a developer did not know how to encrypt/decrypt a binary file. To do so, a parameter use\_binary=true must be passed to the DataSink\_Stream API in the Botan library. In another example, developers discussed how large files can be encrypted by the WebCrypto API.
Developers struggled to use the provided functions in libraries to generate secured random numbers. 

Other discussions were centered on password-based encryption (PBE). 
Developers asked how to configure PBE APIs in libraries such as Jasypt. 
The PBE API commonly requires a password, iteration count, and salt generator, for which developers struggled to assign the correct values.

Lastly, different padding schemes created technical problems for developers. 
Discussions were related to the security level provided between PKCS1.5 and OAEP, the usage of zero padding in OpenSSL, PKCS\#7 padding with AES, and how padding can be disabled in a crypto library.

\boxit{
In the encryption/decryption theme, we found sub-problems in which developers mainly asked for help.
The sub-problems include password-based encryption, paddings, encoding/decoding, modes of encryption, library specific issues, and decryption issues.
}
\begin{table*}
\tiny
\centering
\caption {The number of assigned posts to each theme in a crypto library} \label{tab:alllibs} 
\renewcommand{\arraystretch}{1.5} 
\begin{tabular}{lcccccccccc}
\hline
              & \textbf{\begin{tabular}[c]{@{}c@{}}Encryption\\ Decryption\end{tabular}} & \textbf{\begin{tabular}[c]{@{}c@{}}Library \\ installation\end{tabular}} & \textbf{\begin{tabular}[c]{@{}c@{}}Certificate-related \\ issues\end{tabular}} & \textbf{\begin{tabular}[c]{@{}c@{}}Library \\ interoperability\end{tabular}} & \textbf{\begin{tabular}[c]{@{}c@{}}Generate/store \\ keys\end{tabular}} & \textbf{Hashing}          & \textbf{\begin{tabular}[c]{@{}c@{}}Digital \\ signature\end{tabular}} & \textbf{\begin{tabular}[c]{@{}c@{}}Sample \\ implementation\end{tabular}} & \textbf{\begin{tabular}[c]{@{}c@{}}Random number \\ generator\end{tabular}} & \textbf{\begin{tabular}[c]{@{}c@{}}Cryptographic \\ attacks\end{tabular}} \\ \hline
OpenSSL       & 7                                                                        & 4                                                                        & 6                                                                              & 3                                                                            &                                                                         &                           & 3                                                                     & 1                                                                         &                                                                             & 1                                                                         \\ \hline
Bouncy Castle  & 5                                                                        & 1                                                                        & 8                                                                              & 1                                                                            & 7                                                                       &                           & 2                                                                     & 1                                                                         &                                                                             &                                                                           \\ \hline
CryptoJS      & 7                                                                        & 1                                                                        & 1                                                                              & \cellcolor[HTML]{C0C0C0}13                                                   &                                                                         & 2                         &                                                                       & 1                                                                         &                                                                             &                                                                           \\ \hline
mcrypt        & 5                                                                        & \cellcolor[HTML]{C0C0C0}16                                               &                                                                                & 4                                                                            &                                                                         &                           &                                                                       &                                                                           &                                                                             &                                                                           \\ \hline
PyCrypto      & 6                                                                        & 8                                                                        & 2                                                                              & 6                                                                            &                                                                         &                           & 1                                                                     & 2                                                                         &                                                                             &                                                                           \\ \hline
phpseclib     & 4                                                                        & 7                                                                        & 2                                                                              & 4                                                                            & 2                                                                       &                           & \cellcolor[HTML]{C0C0C0}6                                             &                                                                           &                                                                             &                                                                           \\ \hline
Crypto++      &7                                                                        & 3                                                                       &                                                                                & 2                                                                            & 3                                                                       & 4                         & 4                                                                     & 1                                                                         &                                                                             & 1                                                                         \\ \hline
CryptoAPI     & 6                                                                        &                                                                          & 8                                                                              & 1                                                                            & 4                                                                       & 5                         & 1                                                                     &                                                                           &                                                                             &                                                                           \\ \hline
pyOpenSSL     &                                                                          & 4                                                                        & \cellcolor[HTML]{C0C0C0}17                                                     &                                                                              & 1                                                                       &                           & 3                                                                     &                                                                           &                                                                             &                                                                           \\ \hline
Jasypt        & \cellcolor[HTML]{C0C0C0}9                                                & 11                                                                       &                                                                                & 1                                                                            &                                                                         & 4                         &                                                                       &                                                                           &                                                                             &                                                                           \\ \hline
libsodium     & 8                                                                        & 8                                                                        &                                                                                & 3                                                                            & 2                                                                       & 1                         & 1                                                                     & 1                                                                         & 1                                                                           &                                                                           \\ \hline
M2Crypto      & 2                                                                        & 12                                                                       & 3                                                                              &                                                                              & 3                                                                       &                           & 3                                                                     & 2                                                                         &                                                                             &                                                                           \\ \hline
Web Crypto API & 6                                                                        & 1                                                                        & 2                                                                              & 7                                                                            & 4                                                                       & 2                         & 2                                                                     & 1                                                                         &                                                                             &                                                                           \\ \hline
JCA           & 6                                                                        & 4                                                                        & 4                                                                              &                                                                              & 4                                                                       & 3                         & 3                                                                     & 1                                                                         &                                                                             &                                                                           \\ \hline
CommonCrypto  & 5                                                                        & 5                                                                        &                                                                                & 1                                                                            &                                                                         & \cellcolor[HTML]{C0C0C0}8 & 2                                                                     & \cellcolor[HTML]{C0C0C0}4                                                 &                                                                             &                                                                           \\ \hline
node-crypto   & \cellcolor[HTML]{C0C0C0}9                                                & 1                                                                        & 1                                                                              & 7                                                                            & 3                                                                       & 3                         &                                                                       & 1                                                                         &                                                                             &                                                                           \\ \hline
Botan         & 7                                                                        & 11                                                                       & 2                                                                              & 1                                                                            & 1                                                                       & 2                         &                                                                       &                                                                           & 1                                                                           &                                                                           \\ \hline
Spongy Castle  & 4                                                                        & 5                                                                        & 6                                                                              &                                                                              & \cellcolor[HTML]{C0C0C0}8                                               &                           & 1                                                                     & 1                                                                         &                                                                             &                                                                           \\ \hline
SJCL          & 8                                                                        &                             1                                             &                                                                                & 8                                                                            & 2                                                                       & 3                         &                                                                       & 2                                                                         & 1                                                                           &                                                                           \\ \hline
wolfSSL   & 1                                                                        & 8                                                                        & 12                                                                              & 1                                                                            &                                   1                                      &                           & 2                                                                     &                                                                           &                                                                             &                                                                        \\ \hline
\end{tabular}

\end{table*}
\subsubsection{Library installation}
This theme depicts problems regarding installation, compilation, usage issues, and setting up the prerequisites for a library to work.
This theme has the second-highest number of posts as developer platforms and integrated development environments (IDEs) varied when developers worked with a specific crypto library.
For instance, developers discussed the dependencies of Spongy Castle in the Gradle file, setting up Android Studio with Spongy Castle,  adding JCE to JRE 8 on macOS Sierra, and building the Botan library with the nmake command.
Each crypto library commonly uses a specific way to install or compile of the library or its modules.
For instance, in the PyCrypto and M2Crypto libraries, developers commonly need to resolve their issues with the pip command, and similarly, the usage of npm was the key reason for other discussions related to the node-crypto library. 

\boxit{
It is patently evident that the process of getting a crypto library up and running under different circumstances, \eg platforms or IDEs, can be troublesome for developers.
}

\subsubsection{Certificate-related issues}
We found two sub-problems with the theme of the certificate-related issue.
The first challenge developers encountered was working with various file formats, \eg p7b, and various encodings \eg Privacy Enhanced Mail (PEM) or  Distinguished Encoding Rules (DER).
Developers asked about how to read or save PEM files using crypto libraries, storing/reading private keys in a Public Key Cryptography Standards (PKCS\#8) file, differences between DER and PEM file formats,  and storing/reading public and private keys in a PKCS\#12 file.
The other issue of developers was to extract various elements from a certificate, \eg expiration date, list of Subject Alternative Name (SAN) and Certificate Authority (CA), and cipher list.
They also had challenges in checking a valid certificate, generating a self-signed certificate, using different versions of TLS and SSL, and TLS handshake issues.

\boxit{
Certificates coupled with many cryptographic concepts and this fact complicate working with certificates.
Various PKCS standards and the correct way of establishing a TLS connection are still problematic.
}

\subsubsection{Library interoperability}
It is common for developers to work with more than one crypto library in a large project. 
However, there might be some discrepancies between the libraries. 
A common issue was that developers encrypted a piece of data with OpenSSL, \ie via command line,  and then they had issues with decryption of the ciphertext with another library.
This is due to the fact that the default values in libraries commonly do not match. 
For instance, a developer encrypted a text with OpenSSL but could not decrypt it with the Botan library because of the default usage of PKCS\#1 v1.5 padding in OpenSSL. 
Furthermore, on closer inspection, root causes are mainly the inappropriate encoding of the ciphertext, incorrect IVs, generating cryptographic keys differently, and using unequal key formats and padding options.

\boxit{
Working with more than one library seems to be a challenging task for developers due to different default values in APIs, encodings, paddings, and key generation methods.
}

\subsubsection{Generate/store crypto keys}
For every cryptography scenario, developers need to generate and store their crypto keys.
In the analyzed discussions, the challenges are related to storing keys, \eg AndroidKeyStore, generate a valid ECDSA or RSA key pair,  generate a symmetric key, differences between trust store and keystore, generate keys with Key Based Key Derivation Function (KBKDF), the correct length of possible keys for various algorithms, and the meaning of modulus (n) and public key exponent (e) in RSA keys.

\boxit{
Developers mainly dealt with the differences of crypto keys among symmetric, \eg AES, and asymmetric, \eg RSA or ECDSA, algorithms.
}

\subsubsection{Hashing}
It appears that developers still talk about the possibility of reversing a hashed string. 
However, most of the discussions were about generating a hash string, the right way of using salt, calculating checksum for large files, issues in using Hash-based Message Authentication Code (HMAC), and the usage of hash functions, \ie Password-Based Key Derivation Function 2 (PBKDF2), bcrypt and scrypt.

\boxit{
Compared to the other themes, hashing requires developers to understand fewer concepts and hence, there are fewer discussions in the recent active posts.
}

\subsubsection{Digital signature}
Developers faced issues when signing and verifying a signature. 
A developer misunderstood the application of the Cipher API and the Signature API for signing a piece of data in JCA. Another developer was worried about performance bottleneck when there is a massive dataset. In other discussions, developers failed to verify a signature due to the wrong encoding of RSA keys in browsers (URL encoding), using a hash as data to be signed instead of the data itself, using the wrong key for signing or verification, the mismatched padding for the signature, and verifying a certificate in the chain of trust.

\boxit{
It seems developers suffered from the lack of technical knowledge about digital signatures and issues that are indirect to the topic, \eg browser encoding and default values for padding in crypto libraries.
}


\subsubsection{Sample implementation}
Developers mainly asked for two types of sample implementation. 
In the first type, developers had a sample code from a language or a specific crypto library and were looking for an equivalent piece of code in another language or library.
For instance, a developer had a piece of encryption code in Objective-C but was not able to do the same in Swift.
In the second type, developers had a goal but did not know how the task could be accomplished. 
For example, a developer requested a sample implementation of AES256 CBC in the M2Crypto library.

\boxit{
Documentation of crypto libraries should provide extensive secure examples so that developers have a reliable source of sample implementations. 
}

\subsubsection{Cryptographic attacks}
Only 0.4\% of the analyzed posts were concerned about cyber attacks. 
The first discussion was about conducting a man-in-the-middle attack when a self-signed certificate is used. 
The asker received comprehensive responses regarding why a self-signed certificate is not recommended. 
In the second discussion, a developer was not able to comprehend how the length extension attack works. 

\boxit{Just two discussions explicitly discussed attacks against cryptography. However, such discussions may appear more in crypto.stackexchange.com.
At the same time, most general developers consult \SO as it is more general compared to crypto stackexchange}

We attempted to cast some light on the common technical issues of developers with various crypto libraries. 
We observed that developer uncertainty in a particular crypto library not only is related to one or two areas but is frequently linked to more than five themes.  
There are some libraries, such as OpenSSL and WolfSSL, that are intended to be used for special purposes, \ie secure communications over computer networks. 
This increases the likelihood of identifying more questions related to the certificate issues in such libraries. 
Moreover, a popular crypto library, such as Bouncy Castle, presents a wide range of crypto APIs and be can be utilized in two popular programming languages, \ie C\#, and Java. 
This can explain why identified questions are linked to seven themes.
Some of the extracted themes are interrelated, \eg certificate-related issues, digital signatures, and generating/storing keys. 
For instance, a developer may need to generate an RSA key pair to work with certificates.
 However, we attempted to carefully identify the core issue of the posted challenge. 
 
The detailed issues in working with various crypto libraries could provide valuable support for professionals to identify the probable pitfalls in the design phase of software development. 
Admittedly, identifying crypto pitfalls in earlier stages can substantially boost the security and the speed of development of software.
As a result, such forethoughts can facilitate the use of cryptography in the implementation phase and prevent inexperienced developers from making fatal security mistakes that may have pernicious effects after the release phase. 
Further research is needed to shed light on how similar APIs in popular crypto libraries are misunderstood and whether the complexity of APIs has an impact on creating more problems or not.

\section{Threats to validity}
\label{sec:threat}
We selected 25 posts from each crypto library. 
This may not be a representative sample of the whole population; however, we were particularly interested in the common themes of issues in various libraries, not just one library. 
We selected the latest posts that are active on \SO that had at least one answer and skipped the recent questions to which nobody responded as well as the questions with no positive received votes. 
Nonetheless, there are various approaches to choose the posts, \eg the number of answers or the number of views, while each of them can impose some threats to validity.
To reduce subjectivity, two authors of this paper carefully performed thematic analysis to extract the themes. 
The final list of themes is deduced based on their discussions and cross-checking. 
Nevertheless, a few posts could have been assigned to other themes or a current theme could have been divided into several sub-themes. 
We may not have covered all the crypto libraries discussed on \SO, but we indeed selected the popular ones.

\section{Related work}
\label{sec:related}

Kafader and Ghafari developed FluentCrypto with the goal of creating usable and secure crypto APIs for developers~\cite{kafader2021fluentcrypto}.
FluentCrypto hides the low-level complexities that involve using a native API and provides a task-based solution that novices can use without crypto knowledge.
It also allows crypto experts to configure the API as they find fit and uses a set of pre-defined rules to determine the configuration is secure.
Green \etal proposed ten principles to aid library developers in reducing the possibility of API misuses \cite{green2016developers}. 
For instance, one of the principles is to make defaults safe and unambiguous in APIs. 
This principle can significantly lessen the hardship of developers as we have witnessed their confusion about default values in library interoperability and encryption/decryption themes.
Reviewing 2491 Stack Overflow questions in relation to seven crypto libraries, 
Patnaik \etal identified 16 underlying usability issues in crypto libraries
\cite{patnaik2019usability}. 
There is a common theme between their work and this study, which is the demand for example code snippets. However, their objective was to investigate the usability of crypto APIs, whereas we grouped the problems into themes based on the technical perspective.
Developers refer to \SO as popular documentation.
Parnin \etal studied three popular non-crypto APIs (Android, GWT, Java) to observe the quality and dynamics of the Stack Overflow documentation for these APIs \cite{parnin2012crowd}. 
They found that the crowd generated a rich source of content containing code examples, which are viewed by a majority of developers. For instance, more than 35 000 developers contributed to Android API discussions, which covers 87\% of the classes and has been viewed over 80 million times.
However, with the massive number of discussions, there is a small pool of experts available to answer the questions.
Hou \etal conducted a manual analysis on a set of newsgroup discussions to understand developer problems in using APIs \cite{hou2011obstacles}.
They described 15 obstacles, \eg unclear API semantics by design or wrong parameter values, which hinder developers, and alleviating such obstacles increases the accessibility of APIs.
Hazhirpasand \etal conducted a large-scale study on crypto-related posts on \SO by using Latent Dirichlet Allocation (LDA), which is a generative statistical model, to cluster 91 954 questions \cite{hazhirpasand2021hurdles}. 
They found three high-level themes in developers' questions, namely digital certificate, programming issues, and password/hashing.
In contrast, we reported more elaborated themes of issues specifically related to crypto libraries and excluded configurational problems and general-purpose crypto questions.
A recent study analyzed 489 Java projects in which the majority of APIs (i.e., 13 of 15) were misused at least once \cite{hazhirpasand2020java}. In addition, contacting the developers showed that security warnings in the documentation of crypto APIs are rare. Consequently, based on a recent survey, developers incline to resolve their crypto issues on \SO, where the authenticity of the provided answers is debatable \cite{hazhirpasand2021worrisome}.

\section{Conclusions}
\label{sec:conclusion}
There have been numerous studies to investigate why crypto APIs are hard to use for developers. 
Such studies examined the issues from the developer's point of view as well as the usability of crypto APIs.
We were curious to observe what technical problems are common among different crypto libraries. 
We selected 25 discussions from 20 crypto libraries on \SO and to the best of our knowledge, we did not find any study in which 20 crypto libraries were considered.
We identified 10 themes in the discussions and the majority of libraries were involved in more than five themes. 
There exist 0.04\% of questions concerning attacks against cryptography, whereas  112 questions were related to encryption/decryption issues. 
The developers also asked questions mostly about library installation, digital certificates, crypto keys, and library interoperability.
The implications of these findings can assist security and software professionals to correctly guide their team members when dealing with cryptography, and especially crypto libraries.
Further work is certainly required to disentangle the problematic commonalities among various crypto libraries.


\section{Acknowledgments}
We gratefully acknowledge the financial support of the Swiss National Science Foundation for the project
``Agile Software Assistance'' (SNSF project No.\ 200020-181973, Feb.\ 1, 2019 - April 30, 2022).
We also thank CHOOSE, the Swiss Group for Original and Outside-the-box Software Engineering of the Swiss Informatics Society, for its financial contribution to the presentation of this paper.


\bibliographystyle{IEEEtran}
\bibliography{cryptolib}

\end{document}